\documentclass[journal,comsoc]{IEEEtran}
\usepackage[T1]{fontenc}
\usepackage{caption}
\usepackage{epsfig}
\usepackage{multirow}
\usepackage{enumitem}
\usepackage{hyperref}
\setitemize{noitemsep,topsep=0pt,parsep=0pt,partopsep=0pt}
\usepackage[noend]{algpseudocode}
\usepackage{algorithmicx,algorithm}
\usepackage{enumerate} 
\usepackage{pgfplots}
\newcommand{\etal}{\textit{et al.~}}
\usepackage{graphicx}
\usepackage{pythonhighlight}
\usepackage[english]{babel}
\usepackage[T1]{fontenc}
\usepackage{float}
\usepackage{graphicx}
\usepackage{bm}
\usepackage{braket}
\usepackage{xcolor}
\usepackage{amsfonts}
\usepackage{comment}
\usepackage{dirtytalk}
\usepackage{hyperref}
\usepackage{url}
\usepackage{bbold}
\usepackage{realboxes}

\usepackage{enumitem}
\usepackage{algpseudocode}

\usepackage{listings}
\usepackage{xcolor}
\definecolor{bkgd}{RGB}{240,242,246}
\definecolor{ceruleanblue}{rgb}{0.16, 0.32, 0.75}
\definecolor{orange-red}{rgb}{1.0, 0.27, 0.0}
\definecolor{anotherblue}{RGB}{37,92,243}
\definecolor{blackblue}{RGB}{46,60,85}
\definecolor{goldyellow}{RGB}{199,146,12}
\lstdefinestyle{altstyle2}{
    backgroundcolor=\color{bkgd},
    basicstyle=\ttfamily\footnotesize\color{blackblue},
    breakatwhitespace=false,
    breaklines=true,
    captionpos=b,
    commentstyle=\color{goldyellow},
    keepspaces=true,
    keywordstyle=\color{orange-red},
    language=Python,
    numbersep=5pt,
    numberstyle=\tiny\color{ceruleanblue},
    showspaces=false,
    showstringspaces=false,
    showtabs=false,
    stringstyle=\color{anotherblue},
    tabsize=2
}
\usepackage{amssymb}
\usepackage{bbding}
\usepackage[caption=false]{subfig}
\usepackage{booktabs}
\usepackage{amsmath}
\usepackage{cleveref}
\usepackage[cmintegrals]{newtxmath}
\begin{document}
\title{VQNet 2.0: A New Generation Machine Learning Framework that Unifies Classical and Quantum}
\author{Huanyu Bian, $^{1, 7}$
Zhilong Jia,$^{1,4, 6}$
Menghan Dou, $^{1, 4}$
Yuan Fang, $^{1, 4}$
Lei Li, $^{1,4}$
Yiming Zhao, $^{1, 4}$
Hanchao Wang, $^{1, 4}$
Zhaohui Zhou, $^{1, 4}$
Wei Wang,$^{1, 4}$
Wenyu Zhu,$^{1, 4}$
Ye Li, $^{1, 4}$
Yang Yang, $^{3, 5}$
Weiming Zhang,$^{2,*}$ 
Nenghai Yu, $^{2,*}$
Zhaoyun Chen,$^{5,*}$ 
Guoping Guo$^{5, 6,*}$\\
$^{1}$Origin Quantum Computing Company Limited, Hefei 230026, China\\
$^{2}$CAS Key Laboratory of Electromagnetic Space Information(University of Science and Technology of China), Hefei 230026, China \\
$^{3}$School of Electronics and Information Engineering(Anhui University), Hefei, Anhui 230601, China\\
$^{4}$Anhui Engineering Research Center of Quantum Computing, Hefei 230026, China\\
$^{5}$Institute of Artificial Intelligence( Hefei Comprehensive National Science Center), Hefei, Anhui 230026, China\\
$^{6}$CAS Key Laboratory of Quantum Information (University of Science and Technology of China), Hefei 230026, China\\
$^{7}$Aerospace Information Research Institute, Chinese Academy of Sciences, Beijing 100190, China\\
}
\maketitle
\begin{abstract}
With the rapid development of classical and quantum machine learning, a large number of machine learning frameworks have been proposed. However, existing machine learning frameworks usually only focus on classical or quantum, rather than both. Therefore, based on VQNet 1.0, we further propose VQNet 2.0, a new generation of unified classical and quantum machine learning framework that supports hybrid optimization. The core library of the framework is implemented in C++, and the user level is implemented in Python, and it supports deployment on quantum and classical hardware. In this article, we analyze the development trend of the new generation machine learning framework and introduce the design principles of VQNet 2.0 in detail: unity, practicality, efficiency, and compatibility, as well as full particulars of implementation. We illustrate the functions of VQNet 2.0 through several basic applications, including classical convolutional neural networks, quantum autoencoders, hybrid classical-quantum networks, etc. After that, through extensive experiments, we demonstrate that the operation speed of VQNet 2.0 is higher than the comparison method. Finally, through extensive experiments, we demonstrate that VQNet 2.0 
can deploy on different hardware platforms, the overall calculation speed is faster than the comparison method. It also can be mixed and optimized with quantum circuits composed of multiple quantum computing libraries.    
\end{abstract}
\begin{IEEEkeywords}
Machine Learning Framework; Quantum Machine Learning; Classical Machine Learning
\end{IEEEkeywords}
\IEEEpeerreviewmaketitle
\section{Introduction}
With the rapid development of quantum computing, quantum machine learning, which combines quantum computing with classical machine learning, has also developed rapidly~\cite{huang2021power, guan2021quantum, lamata2020quantum, gyongyosi2020optimizing,biamonte2017quantum, ciliberto2018quantum, von2018quantum, christensen2020fchl}. Quantum computing is a new computing method that uses quantum properties such as superposition and entangled states to perform calculations. Compared with classical computing, quantum computing has two advantages: 1) Theoretically, quantum computers far surpass classical computers in calculation speed and storage efficiency. 2) With the continuous size reduction of classical electronic components, the development of classical computers is about to face the "Size Effect" and other issues that hinder continued rapid developments. Quantum machine learning will further improve the processing capabilities of big data by using the computational efficiency of quantum computers far surpassing classical computers, combined with machine learning algorithms that have developed rapidly in the era of big data.

\begin{table}[t]
\scriptsize
\renewcommand\arraystretch{1.2}
\caption{Classification of machine learning frameworks.}
\label{tab:categories}
\begin{center}
\begin{tabular}{c|c|c}
\hline
      & Classical machine learning & Quantum machine learning  \\
\hline\hline
Caffe & \Checkmark &  \XSolid  \\
\hline
TensorFlow & \Checkmark &  \XSolid \\
\hline
Theano & \Checkmark &  \XSolid \\
\hline
PyTorch & \Checkmark &  \XSolid \\
\hline\hline
TensorFlow Quantum & \XSolid &  \Checkmark \\
\hline
PennyLane & \XSolid &  \Checkmark \\
\hline
Paddle Quantum & \XSolid &  \Checkmark \\
\hline\hline
VQNet 2.0 & \Checkmark &  \Checkmark \\

\hline
\end{tabular}
\end{center}

\end{table}

Meanwhile, the increasing interest in classical machine learning and quantum machine learning has led to an emergence of quantum machine learning frameworks. Although many previous works have developed a variety of machine learning frameworks, there is still a lack of a new generation of machine learning frameworks to support both classical machine learning and quantum machine learning, as shown in Tab.~\ref{tab:categories}. To solve the above-mentioned lack of the new generation machine learning framework, based on VQNet 1.0~\cite{chen2019vqnet}, we propose VQNet 2.0, a Python framework that unifies classical machine learning and quantum machine learning, which means 1) unifying classical and quantum machine learning algorithms, and 2) unifying classical and quantum computers. Its software stack is shown in Fig.\ref{fig:abstract_pip}. In addition to the above unification of machine learning algorithms, VQNet 2.0 also achieves the unification of classical computers and quantum computers: 1) Classical nodes are deployed on classical computers, similar to classical machine learning frameworks. 2) We design Qpanda~\cite{menghan2022QPanda}, a software package that connects hardware and quantum computing. Through Qpanda, quantum nodes are deployed on the quantum processing unit (QPU) or quantum simulator.

\begin{figure}[!t]
\centering
\includegraphics[width=0.99\linewidth]{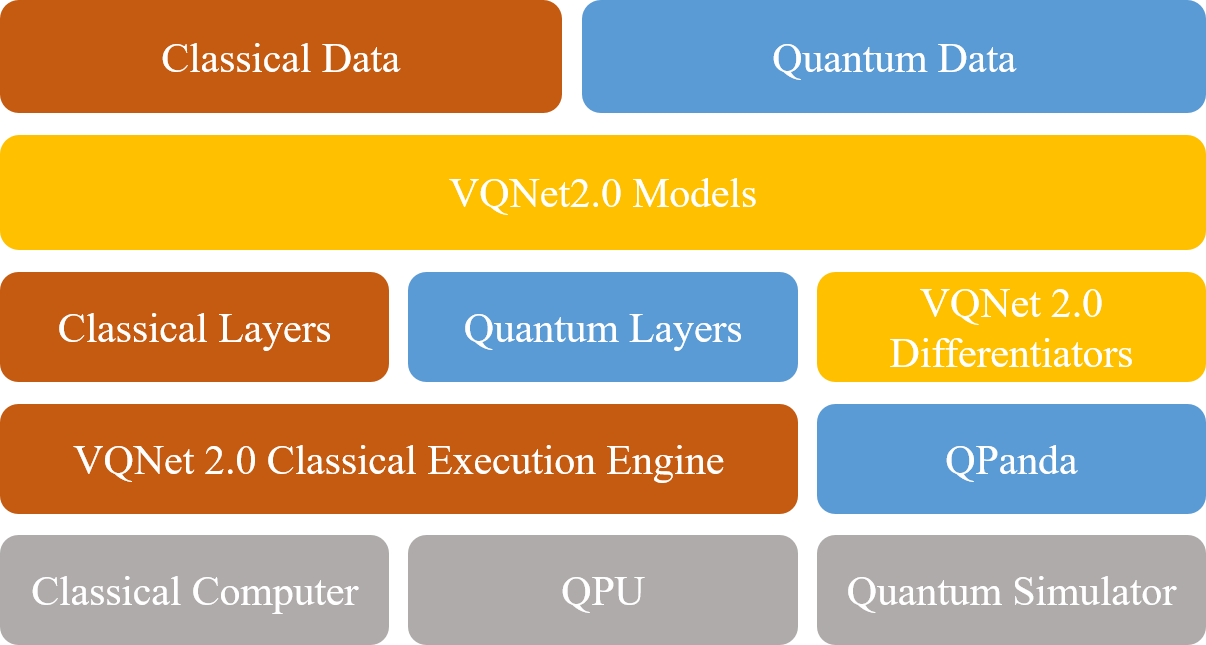}
\vspace{0.8em}
\caption{VQNet 2.0's software stack. At the top of the stack is the data to be processed. VQNet 2.0 has the ability to process classical data (images, text, etc.) and quantum data (quantum circuits, quantum operators, etc.). Next is the model API of VQNet 2.0. Then comes the computing layers and the differentiators. The computing layers include the classical layers and the quantum layers, and the differentiators can perform hybrid automatic differentiation for networks built on any computing layer. After that, the quantum circuit is accomplished by QPanda, and the entire network can be operated on a classical computer, quantum simulator, or QPU according to its function.}
\label{fig:abstract_pip}
\end{figure}

The rest of the paper is organized as follows. 
In Section II, we describe the background of our proposed framework and the trend of the machine learning framework. 
In Section III, we describe our design principles.
In Section IV, we give the pipeline of our VQNet 2.0, which can support both classical machine learning and quantum machine learning. 
Section V shows the implementation details of our proposed framework.
Section VI and Section VII give the extensibility of case studies which contain both classical and quantum applications. 
Section VIII shows the experimental setup and experimental results of our VQNet 2.0 compared with other state-of-the-art frameworks. 
Section IX draws the conclusion.

\section{Background}
\subsection{VQNet 1.0}
Quantum computing is a completely new method that can revolutionize computers. In order to explore the possibility of hybrid implementation of the variable quantum algorithm and classical machine learning framework, Chen \etal~\cite{chen2019vqnet} proposed VQNet 1.0, a quantum-classical hybrid machine learning architecture. VQNet 1.0 designed two kinds of  quantum operators (QOPs): \Colorbox{bkgd}{\lstinline{qop}} and \Colorbox{bkgd}{\lstinline{qop_pmeasure}}. Through making these two QOPs support forward and backward propagation at the same time, VQNet 1.0 
could achieve model training. Furthermore, it is possible to use a gradient-based optimizer for network optimization in VQNet 1.0. Chen \etal used VQNet 1.0 to construct a variety of quantum machine learning algorithms, including QAOA, VQE algorithm, quantum classifier, and quantum circuit learning, and they all achieved good performance.

Although VQNet 1.0 effectively connected machine learning and quantum algorithms, it only supports a few types of quantum machine learning algorithms like QAOA, VQE algorithm, quantum
classifier and quantum circuit learning. With the rapid development of machine learning algorithms, it is urgent to propose a new generation of machine learning frameworks based on VQNet 1.0.

\subsection{Trend of Machine Learning Framework}
Compared with the previous single-class machine learning frameworks, we summarize the development trends of the next-generation machine learning frameworks:

First of all, classical machine learning become popular in recent years, and the boom period of quantum machine learning begins. So far, a large number of classical machine learning frameworks, such as Caffe\cite{jia2014caffe}, TensorFlow\cite{abadi2016tensorflow}, Theano\cite{schreiber2017pomegranate}, PyTorch\cite{paszke2019pytorch}, and quantum machine learning frameworks, such as TensorFlow Quantum\cite{broughton2020tensorflow}, PennyLane\cite{bergholm2018pennylane}, Paddle Quantum\cite{Paddlequantum}, have emerged. These frameworks meet the needs of researchers and industrial production for mature frameworks, lower the threshold for research, and further promote the development of classical quantum machine learning. However, with the maturity of classical machine learning algorithms and the continuous progress of quantum machine learning, quantum machine learning will inevitably become a supplement to classical machine learning. The interaction of these two kinds of algorithms make researchers and developers an urgent need for a new generation of machine learning framework that can \textbf{unify classical and quantum}.


Secondly, various gradient-based optimization methods play an extremely important role in the training and convergence of classical deep learning models. Because automatic differentiation\cite{baydin2018automatic} greatly reduces the programming difficulty of classical machine learning, it has become the core software design paradigm of today's machine learning frameworks. However, the existing automatic optimization\cite{paszke2017automatic} algorithms are difficult to directly apply to the training of quantum machine learning models: the gradients of parameterized quantum circuits in quantum machine learning must be evaluated using quantum algorithms, so compared to the gradients of classical machine learning, the existing automatic differentiation algorithm may not work. One type of algorithm in quantum machine learning is hybrid quantum-classical networks, which requires quantum computing and classical computing to cooperate with each other, so \textbf{hybrid optimization}\cite{bergholm2018pennylane} that can output hybrid calculation results is the core function of the new machine learning framework design.

Third, Python\cite{van1995python, oliphant2007python} is an object-oriented, interpreted, general, and open source scripting programming language. It has the characteristics of simple and easy to use, low learning cost, and high writing efficiency, but it has the corresponding disadvantage of slower running speed. Python has been used in a large number of machine learning frameworks such as TensorFlow, PyTorch, etc. Meanwhile, C++\cite{meyers2005effective, alexandrescu2001modern} is an object-oriented compiled language, which is convenient for the direct operation of the memory. C++ not only has the practical characteristics of the efficient operation of computers, but also is committed to improving the programming quality of large-scale programs and the problem description ability of programming languages. It is often used in industrial applications for image processing. Correspondingly, it has shortcomings such as high code complexity and being unfriendly to users. Therefore, in order to avoid these two language shortcomings while inheriting the advantages of both, it is inevitable to \textbf{separate the user level in various languages from the core library}\cite{abadi2016tensorflow}. The core library is implemented in C++, which improves operating efficiency and portability; while the user-level implementation is implemented in Python, which improves user-friendliness and reduces learning costs and difficulties.

Finally, in order to use general-purpose massively parallel hardware (such as GPU) for computational acceleration, existing machine learning frameworks can be deployed on classical hardware such as CPU, GPU, and TPU\cite{collobert2011torch7, chetlur2014cudnn, lavin2015maxdnn}. The rapid increase in the computing power of quantum computers has provided a solid foundation for the development of quantum machine learning. At the same time, the current shortage of quantum computing power makes quantum simulation on classical computers the first choice for most researchers to conduct verification experiments. Therefore, it is inevitable that the new generation of machine learning frameworks can be \textbf{deployed on more types of hardware} such as classical computer, QPU, and quantum simulator.

Based on the above trends, VQNet 2.0 provides a unified classical and quantum machine learning framework that supports hybrid optimization. The core library of the framework is implemented in C++, the user-level is implemented in Python, and it supports deployment on both quantum and classical hardware.

\section{Design Principles}

There are four main principles behind the design of VQNet 2.0:
\subsection{Be United}
VQNet 2.0 aims to achieve multiple unities of classical and quantum: 1) The unification of classical machine learning and quantum machine learning, that is, compared with frameworks such as Qiskit and Pennylane, VQNet 2.0 does not need to import Torch, Tensorflow, or other huge classical machine learning libraries. VQNet 2.0 comes with a complete neural network module, including classical and quantum machine learning, optimizer, loss function, etc. It is easy to quickly build hybrid quantum-classical machine learning models. 2) The deployment of classical computers and quantum computers is unified, that is, compared with frameworks such as TensorFlow and TensorFlow quantum, VQNet 2.0 can deploy classical machine learning models on classical computers, and deploy quantum machine learning models on QPU and quantum simulators.

\subsection{Put Practicality First}
VQNet 2.0 strives to use the framework to build models, train networks, and deploy multiple machines as simple and efficient as possible. In order to bring maximum practicality, it is necessary to realize the following functions including but not limited to: 1) Friendly interface. VQNet 2.0 implements a wealth of quantum machine learning models and interfaces for quickly building quantum machine learning models. VQNet 2.0 performs well in constructing these quantum machine learning algorithms. The interface definition is similar to the syntax of commonly used machine learning frameworks, and users can quickly get started. 2) Automatic differentiation. The realization of automatic differentiation can help VQNet 2.0's users get rid of tedious mathematical derivation and code, further improving the degree of automation and efficiency of using VQNet 2.0. 3) 

Dynamic computation graph. VQNet 2.0 uses the dynamic computation graph mechanism, that is, the construction and calculation of the calculation graph are performed at the same time, which reduces the difficulty of debugging and which also improves the friendliness of programming.

\subsection{Provide Sufficient Performance}
As a general machine learning framework, VQNet 2.0 needs to provide sufficient performance, but not at the expense of practicality. Compared with quantum machine learning frameworks such as Paddle Quantum, Mindspore\cite{Mindspore}, etc., VQNet 2.0 uses QPanda as the quantum computing module, it is possible to build more fully functional and more efficient quantum circuits on classical and quantum computers, and achieve faster forward and 
backward speed.
\subsection{A Good Framework Embraces the Rest}

As a new generation of machine learning framework, VQNet 2.0 is designed to be compatible with other quantum computing libraries. In order to maximize compatibility with Python-based frameworks that can have any configuration system, VQNet 2.0 is designed with compatibility: A compatibility layer is designed to call other frameworks, which is convenient for users to take the quantum circuits described by other frameworks as the black box for optimization calculation, facilitate cross model cooperation, and is suitable for different business and research applications.

\section{Pipeline of VQNet 2.0}
Here, we give an abstract overview of the computational steps involved in the general training and testing of machine learning models in VQNet 2.0.

\begin{figure*}[!t]
\centering
\includegraphics[width=.8\linewidth]{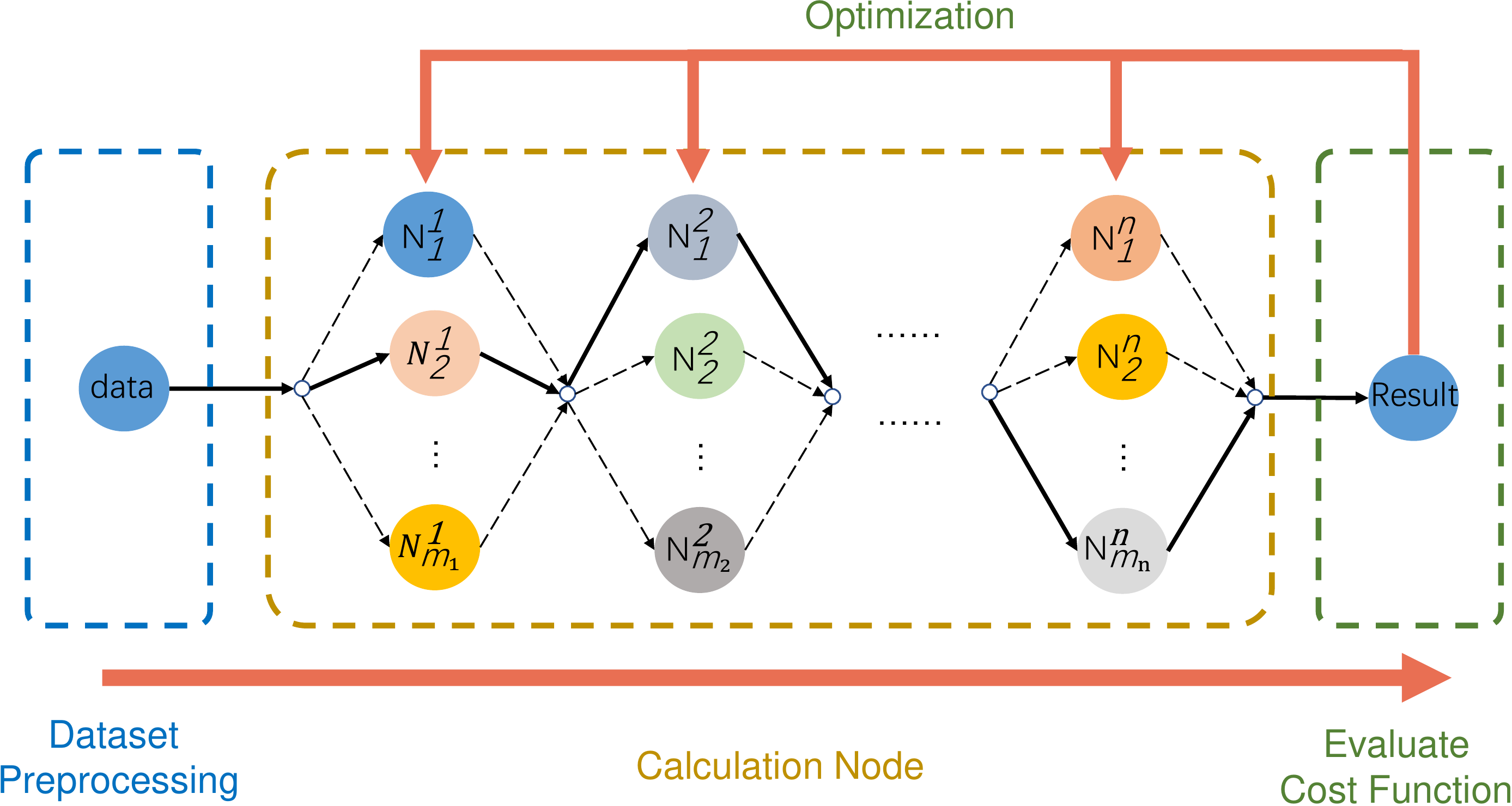}
\vspace{0.3em}
\caption{Abstract pipeline for VQNet 2.0.}
\label{fig:abstract_pip}
\end{figure*}

\subsection{Dataset Preprocessing}
\subsubsection{Classical Data}
Usually, classical data comes from Internet or user-built datasets. In VQNet 2.0, it is easy to download commonly used public data sets by using the functions in \Colorbox{bkgd}{\lstinline{pyvqnet.data}}, similar to the operations in PyTorch and TensorFlow.

\subsubsection{Quantum Data}
Since it is currently impossible to import quantum data from the outside into a QPU, users need to specify the quantum circuit that generates quantum data, while encoding the classical data into the quantum state. There are mainly two existing encoding methods: 1) Encoding features into the basic state, which means that after the classical data is converted into integer form, the binary expansion is expressed as a direct product state, and this encoding method can be realized by using function \Colorbox{bkgd}{\lstinline{pyvqnet.qnn.template.BasicEmbeddingCircuit}}
.2) Encoding features into the rotation angles, which refers to the rotation angle of encoding classical data into a qubit, and this encoding method can be realized by using the function \Colorbox{bkgd}{\lstinline{pyvqnet.qnn.template.AngleEmbeddingCircuit}}. 3) Encoding features into the amplitude vector, which refers to the amplitude vector of converting classical data into a quantum state through a quantum gate operation, and this encoding method can be realized by using the function \Colorbox{bkgd}{\lstinline{pyvqnet.qnn.template.AmplitudeEmbeddingCircuit}}.

\subsection{Calculation Node}
After the data is processed, it can be sent to calculation nodes. In VQNet 2.0, calculation nodes include classical nodes and quantum nodes. In the network built by VQNet 2.0, any number of classical nodes and quantum nodes can be used to connect in any way. It is worth noting that if the number of quantum nodes selected is 0, the network built is a classical machine learning model, and if the number of classical nodes and quantum nodes is not 0, it is a hybrid quantum-classical network.

\subsubsection{Classical Node}
Classical node refers to various network components that process classical data. VQNet 2.0 includes a variety of common basic building blocks, including convolutional layers\Colorbox{bkgd}{\lstinline{pyvqnet.nn.conv}}, pooling layers\Colorbox{bkgd}{\lstinline{pyvqnet.nn.pooling}}, and normlization layers\Colorbox{bkgd}{\lstinline{pyvqnet.nn.batch_norm}}, etc. Various layers can extract features of different dimensions from the original data, and the combination of different features can finally achieve amazing performance.
\subsubsection{Quantum Node}
The quantum node is responsible for calculating quantum data. \Colorbox{bkgd}{\lstinline{QuantumLayer}} simulate a parameterized quantum circuit and get the measurement result. It inherits from \Colorbox{bkgd}{\lstinline{Module}}, so that it can calculate gradients of circuits parameters, and trains variational quantum circuits model or embeds variational quantum circuits into hybrid quantum and classical model. The user defines the function \Colorbox{bkgd}{\lstinline{qprog_with_meansure}} as a parameter, it needs to include the quantum circuit defined by QPanda: generally includes the quantum circuit's coding circuit, evolution circuit, and measurement operation. This class can be embedded in the hybrid quantum-classical machine learning model, and the objective function or loss function of the hybrid quantum-classical model can be minimized through the gradient descent method.

It is particularly worth noting that, in addition to the different calculations, there is another huge difference between quantum nodes and classical nodes. If quantum nodes are required to output classical data, measurement operations are required.
Quantum measurement refers to the interference of the outside world to the quantum system to obtain the required information. The measurement gate uses the Monte Carlo method of measurement. After the qubit is measured, the measurement result will be stored in the classical register. VQNet 2.0 uses \Colorbox{bkgd}{\lstinline{pyvqnet.qnn.measure}} for measurement operations.

\subsection{Loss Function}

After passing through the calculation node, the loss function needs to be calculated next. The loss function selection depends on the completion of the task. It is especially important to note that the calculation of the existing loss function needs to be performed on a classical computer but not a quantum device, so if the network is built with quantum nodes as the last network layer, it must be measured before the loss calculation. VQNet 2.0 builds common loss functions in class \Colorbox{bkgd}{\lstinline{pyvqnet.nn.loss}}.

\subsection{Optimization}

After setting the loss function, the network parameters are updated through back propagation, that is, model training is performed. Common optimization methods are mostly based on gradient descent. VQNet 2.0 constructs common optimization strategies in the class \Colorbox{bkgd}{\lstinline{pyvqnet.optim.optimizer}}. As mentioned earlier, if there is no quantum node in the model involved in the calculation, automatic differentiation can be directly used to calculate the gradient and perform back propagation. If the model is a hybrid quantum-classical network, the hybrid optimization strategy\cite{bergholm2018pennylane} needs to be used to train the network.

\section{Implementation of VQNet 2.0}

\subsection{Be United}
\subsubsection{Unification of Classical and Quantum Machine Learning}

In order to achieve the unity of classical and quantum in one machine learning framework, VQNet 2.0 defines an abstract class \Colorbox{bkgd}{\lstinline{pyvqnet.nn.module.Module}}, which is the base class of all neural network modules. All calculation nodes are designed to be sub-class of class \Colorbox{bkgd}{\lstinline{pyvqnet.nn.module.Module}}, so that they can be used for automatic differential calculation.
\Colorbox{bkgd}{\lstinline{Module}} can also contain other Modules, allowing to nest them in a tree structure. 
In VQNet 2.0, it is easy to assign the sub-modules as regular attributes:

\begin{lstlisting}[language=Python]
class Model(Module):
    def __init__(self):
        super(Model, self).__init__()
        self.conv1 = pyvqnet.nn.Conv2d(1, 20, (5,5))
        self.conv2 = pyvqnet.nn.Conv2d(20, 20, (5,5))

    def forward(self, x):
        x = pyvqnet.nn.activation.relu(self.conv1(x))
        return pyvqnet.nn.activation.relu(self.conv2(x))
\end{lstlisting}        

Sub-modules assigned in this way will be registered.

\subsubsection{Unification of Classical and Quantum Machine}
VQNet 2.0 not only supports the unification of quantum and classical machine learning but also supports operations on classical computers and quantum computers. It should be noted that different calculation nodes have different hardware requirements: 1) For classical nodes, the tools provided by VQNet 2.0 can be directly operated on the classical computer. 2) For quantum nodes, VQNet 2.0 can accomplish real quantum hardware interaction and quantum simulator interaction through QPanda.

For the classical node, since the calculation of classical modules can only be implemented on the classical computer, there is no need for additional specifications in VQNet 2.0. By default, all classical nodes are run in a classical computer and the gradient is obtained through classical methods.

Due to the current lack of quantum computer resources, quantum simulator that simulates the operating environment of QPU on classical computers has become one choice of most scientific researchers. In order to improve the general capabilities of the framework, for quantum node, VQNet 2.0 defines a class \Colorbox{bkgd}{\lstinline{QuantumLayer}}. Users can set the parameter \Colorbox{bkgd}{\lstinline{machine_type}} in \Colorbox{bkgd}{\lstinline{QuantumLayer}} according to their own computing resources and call the corresponding interface provided by QPanda to use QPU or quantum simulator to conduct quantum calculations. Because different quantum simulators have different simulation effects, VQNet 2.0 provides five different kinds of simulators to realize different simulation environments, namely full-amplitude quantum simulator, single-amplitude quantum simulator, partial-amplitude quantum simulator, tensor network quantum simulator, and noise inclusive quantum simulator.

At the same time, in order to enable researchers to better simulate the running results of QPU on the simulator when the calculation resources of QPU are short, VQNet 2.0 adds the noise of QPU to the quantum simulator. VQNet 2.0 adds quantum mapping based on the topology between QPU's qubits to the noise simulator, so as to simulate more real experimental results. The simulation of the noise-containing quantum simulator is closer to QPU. VQNet 2.0 can customize the supported logic gate types and customize the noise model supported by the logic gates. The existing supported quantum noise models are defined in QPanda. VQNet 2.0 uses \Colorbox{bkgd}{\lstinline{NoiseQuantumLayer}} to define an automatic micro-classification of QPU. The user defines function \Colorbox{bkgd}{\lstinline{qprog_with_measure}} as the parameter, which needs to include the quantum circuit defined by QPanda, and also needs to pass in the parameter \Colorbox{bkgd}{\lstinline{noise_set_config}}, using the QPanda interface to set the noise model.

\subsection{Practicality Centric Design}
\subsubsection{Friendly Interface}
The core element to achieve interface friendliness is to design for the interface. The friendliness of the interface definition not only facilitates the maintenance of VQNet 2.0 itself, but also avoids excessive changes to the upper application layer, reducing the cost and difficulty of users. First, in order to provide a more convenient and concise interface, VQNet 2.0 shields the coordination details of multiple interfaces, and provides a coarse-grained interface for upper-layer applications. That is,  the user only needs to define the quantum circuit and network structure. After constructing the forward propagation path, VQNet 2.0 can automatically constitute the back propagation and automatic optimization. Secondly, VQNet 2.0 follows the conventions of most frameworks and uses scene-oriented interface naming methods, such as naming the class that performs one-dimensional convolution operations as class \Colorbox{bkgd}{\lstinline{Conv1D}} to further enhance the friendliness of the interface. Finally, for interface parameters, VQNet 2.0 implements the principle of minimum granularity, that is, to ensure that all parameters passed are used by the interface.

\subsubsection{Dynamic Computation Graph}

Similar to PyTorch, VQNet 2.0 also uses the dynamic computation graph. Computation graphs are directed acyclic computation graphs used to describe operations. They are composed of two main elements: Node which represents data, and Edge which represents operations. In VQNet 2.0, predefined classical and quantum nodes, such as convolutional layer and quantum layer, are also nodes in the computation graph. Edge is the operation performed by each node, \Colorbox{bkgd}{\lstinline{QTensor}} is the basis for constructing the computation graph and the computation graph forms the structural basis of forward and backward propagation. Dynamic graph means that the operation and construction of machine learning model are established at the same time, that is, the value of nodes can be calculated according to the order of forward propagation, and then the computation graph is constructed on the basis of these nodes, and at last the graph constructed by the classical and quantum nodes is used to get the gradient of \Colorbox{bkgd}{\lstinline{QTensor}}. The advantage of the dynamic computation graph is that it is flexible and easy to debug.

\subsection{Implementation with Fabulous Performance}
How to implement an efficient machine learning framework has always been a great challenge for researchers and developers. In order to improve the efficiency of the new generation machine learning framework, in addition to optimizing classical algorithms and classical computers, two aspects need to be considered: 1) the interaction between classical and quantum machine learning algorithms, and 2) the interaction between classical and quantum computers. In order to solve the above problems, VQNet 2.0 makes improvements in two aspects: 1) open-source \textbf{QPanda}, which is a software package specially designed to perform QPU operations and also simulate quantum circuits on classical computers. 2) \textbf{
Unified structure.} The classical and quantum nodes are unified in design.

\textbf{QPanda.} Three high-performance simulations are proposed in QPanda, namely 1) OpenMP multi-thread optimization. Using OpenMP instructions to execute functions in parallel can effectively execute the iterations in the function and effectively reduce the calculation time. In QPanda, the backends of all simulators running large quantum circuit calculations are parallelized to maximize the workload of each core and minimize the overhead caused by multithreading. 2) GPU optimization. GPU has higher memory bandwidth than CPU. Because GPU has a large number of processing cores, it also has higher peak performance than CPU. In QPanda, when the GPU is detected, the GPU is used to accelerate the quantum simulation. 3) Supercomputer simulation. Supercomputers represent the peak of classical computing power in the world, and quantum computing simulations are based on classical computers. QPanda proposes to distribute tasks on multiple calculation nodes on the supercomputer platform ("Shenwei Light of Taihu Lake"). The operation control core runs the program instruction set one by one and calls the operation core to perform core computing tasks, that is, to improve computing efficiency through a two-level parallel computing method.

\textbf{Unified structure.} 1)Unified data structure. Similar to machine learning frameworks such as Torch and Paddle, in VQNet 2.0, the data structure is uniformly defined as \Colorbox{bkgd}{\lstinline{QTensor}}. \Colorbox{bkgd}{\lstinline{QTensor}} is the basic data structure of quantum circuits and neural networks. It supports multiple operations and realizes their derivative functions. When performing various operations required for quantum and classical machine learning calculations, data can be stored by creating \Colorbox{bkgd}{\lstinline{QTensor}}. It can be created using the interface of class \Colorbox{bkgd}{\lstinline{Creation}}. In addition, \Colorbox{bkgd}{\lstinline{QTensor}} can use Numpy, a popular matrix scientific computing package, to load data conveniently.
2) Unified class inheritance. In VQNet 2.0, the classes of quantum and classical machine learning models need to be inherited from the abstract class \Colorbox{bkgd}{\lstinline{Module}} to perform autograd calculations.

\subsection{A Good Framework Embraces the Rest}
VQNet 2.0 hybrid quantum-classical optimization can support not only the quantum circuits of QPanda but also the quantum circuits composed of other quantum computing frameworks (such as Cirq, Qiskit, etc.). VQNet 2.0 provides a quantum circuit computing interface \Colorbox{bkgd}{\lstinline{Compatiblelayer}} for automatic differentiation. The parameters of \Colorbox{bkgd}{\lstinline{Compatiblelayer}} need to be passed in a class, which defines the third-party library quantum circuit, as well as its operation and measurement function \Colorbox{bkgd}{\lstinline{run}}. By using \Colorbox{bkgd}{\lstinline{Compatiblelayer}}, the input of quantum circuits and automatic differentiation of parameters can be implemented by VQNet 2.0.
Class \Colorbox{bkgd}{\lstinline{Compatiblelayer}} is an abstract wrapper to use other framework’s quantum circuits(such as \Colorbox{bkgd}{\lstinline{qiskit.QuantumCircuit}} in Qiskit, \Colorbox{bkgd}{\lstinline{cirq.Circuit}} in Cirq) to forward and backward in the form of VQNet 2.0. The users need to define quantum circuits in the \Colorbox{bkgd}{\lstinline{forward()}} and \Colorbox{bkgd}{\lstinline{backward()}} functions.

\section{Evaluation}

In this section, we first introduce the experimental settings, and then give the deployment results of VQNet 2.0 on different hardware to support its heterogeneous capabilities. After that, we support its efficiency by comparing computing efficiency with different frameworks under different conditions. Finally, an example of VQNet 2.0's support for Qiskit is given to show the compatibility of the framework.

\subsection{Experiment Setup}
\noindent\textbf{Dataset.} For all subsequent experiments, we use the common MNIST data set in computer vision as the experimental data set. The MNIST data set is the handwritten character library. There are 70,000 images, of which 60,000 are the training set and 10,000 are the test set.

\noindent\textbf{Model.} For classical machine learning deployment, we use the most popular neural network: convolutional neural networks(CNN). The CNN model consists of two convolution layers and one fully connected layer. Each convolution layer performs the max-pooling operation after the relu activation function. For quantum machine learning deployment, we use quantum autoencoder for efficient compression of quantum data as pure quantum networks, and
hybrid quantum-classical neural networks as hybrid networks. To verify the compatibility of VQNet 2.0, we use IBM’s
Qiskit quantum computing library for quantum machine learning tasks. All running instances are shown in the Appendix.

\subsection{Deployment}

In order to show that our proposed framework can be deployed normally on different hardware devices, the following shows the deployment results of VQNet 2.0 on the classical computer and QPU.

\subsubsection{Classical Machine Learning Deployment}

\begin{figure}
\captionsetup[subfloat]{font=small, labelformat=simple}
\centering

\subfloat[]{\includegraphics[width = 200pt]{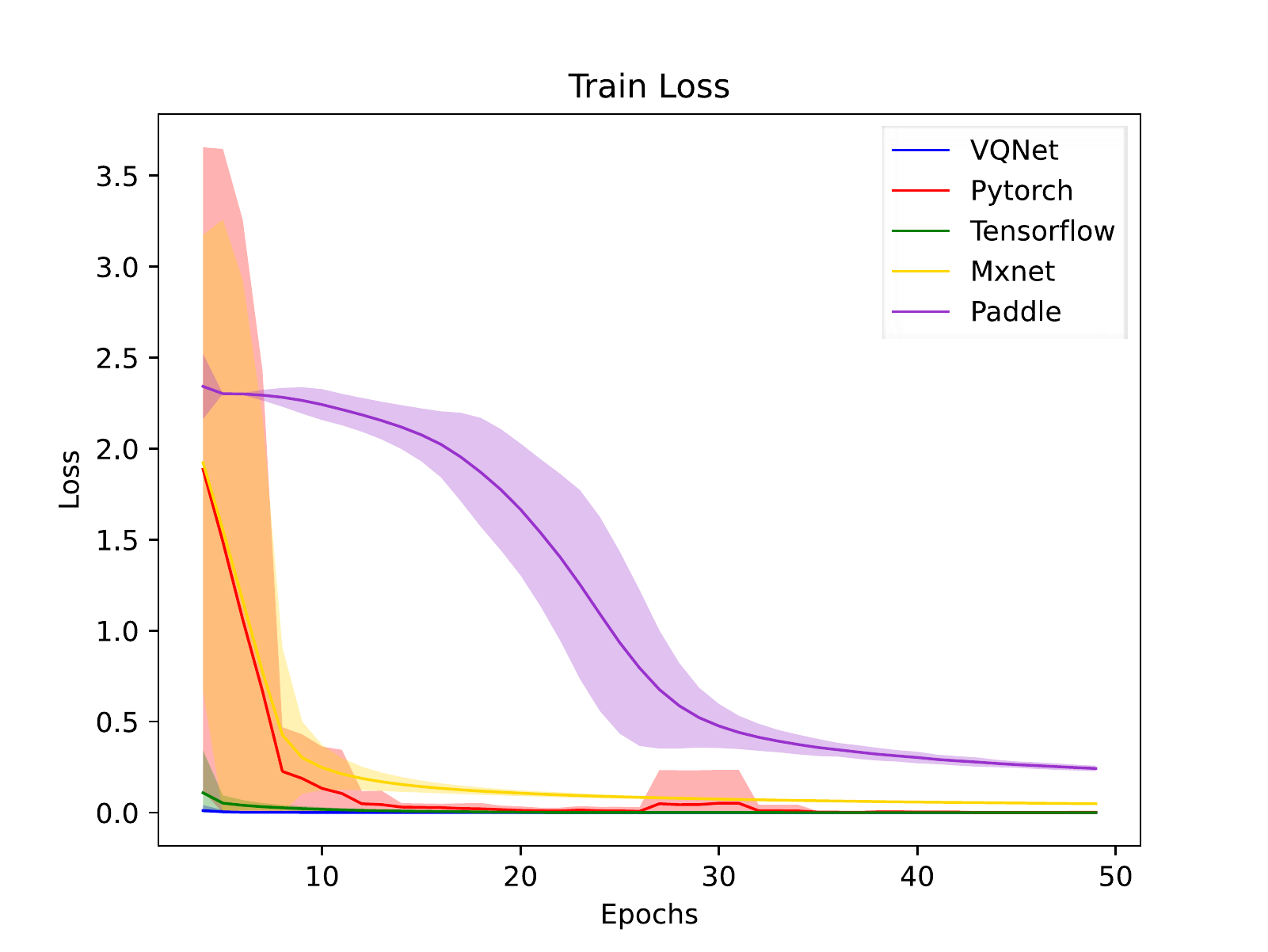}}\\
\subfloat[]{\includegraphics[width = 200pt]{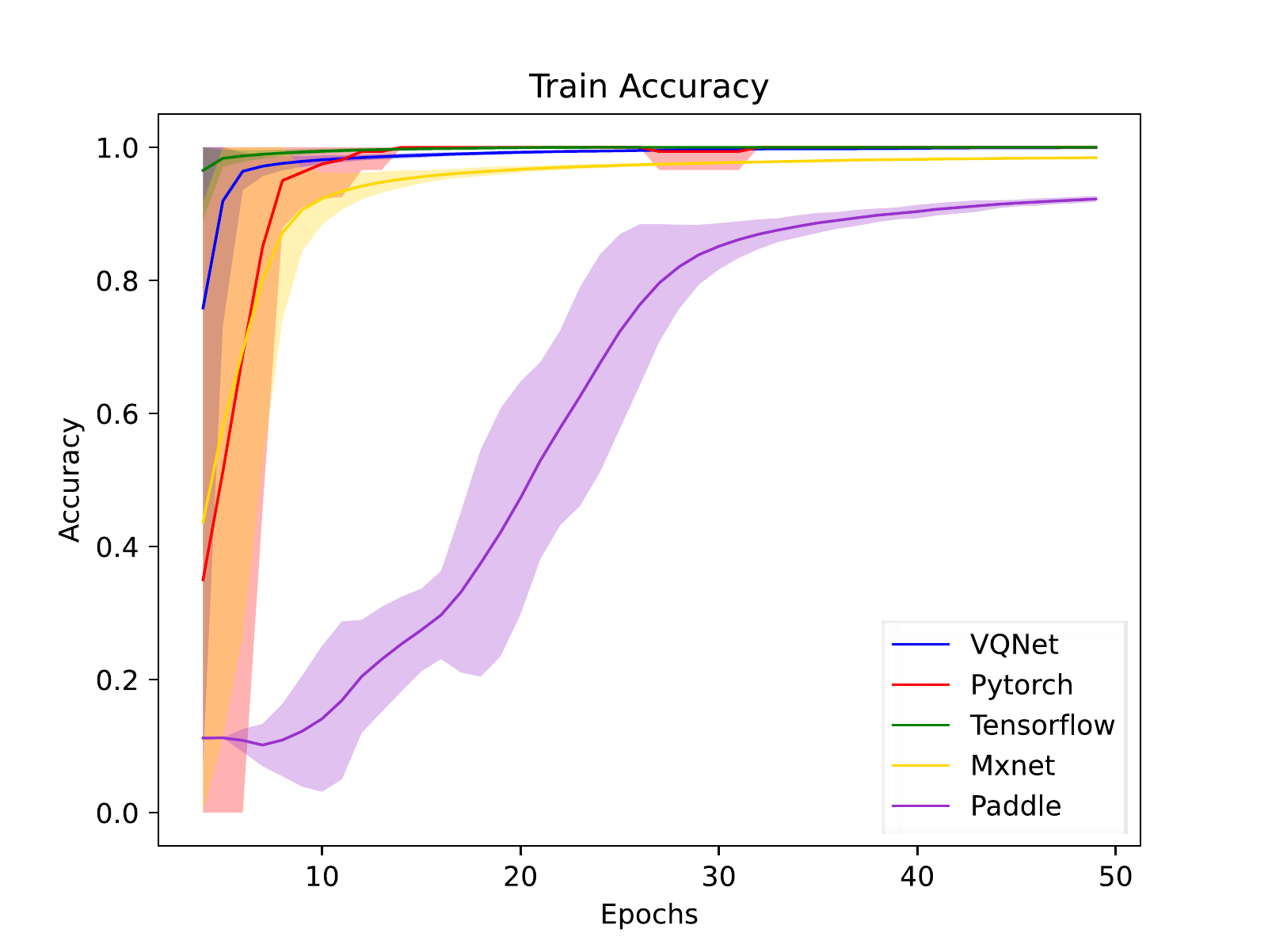}}\\
\subfloat[]{\includegraphics[width = 200pt]{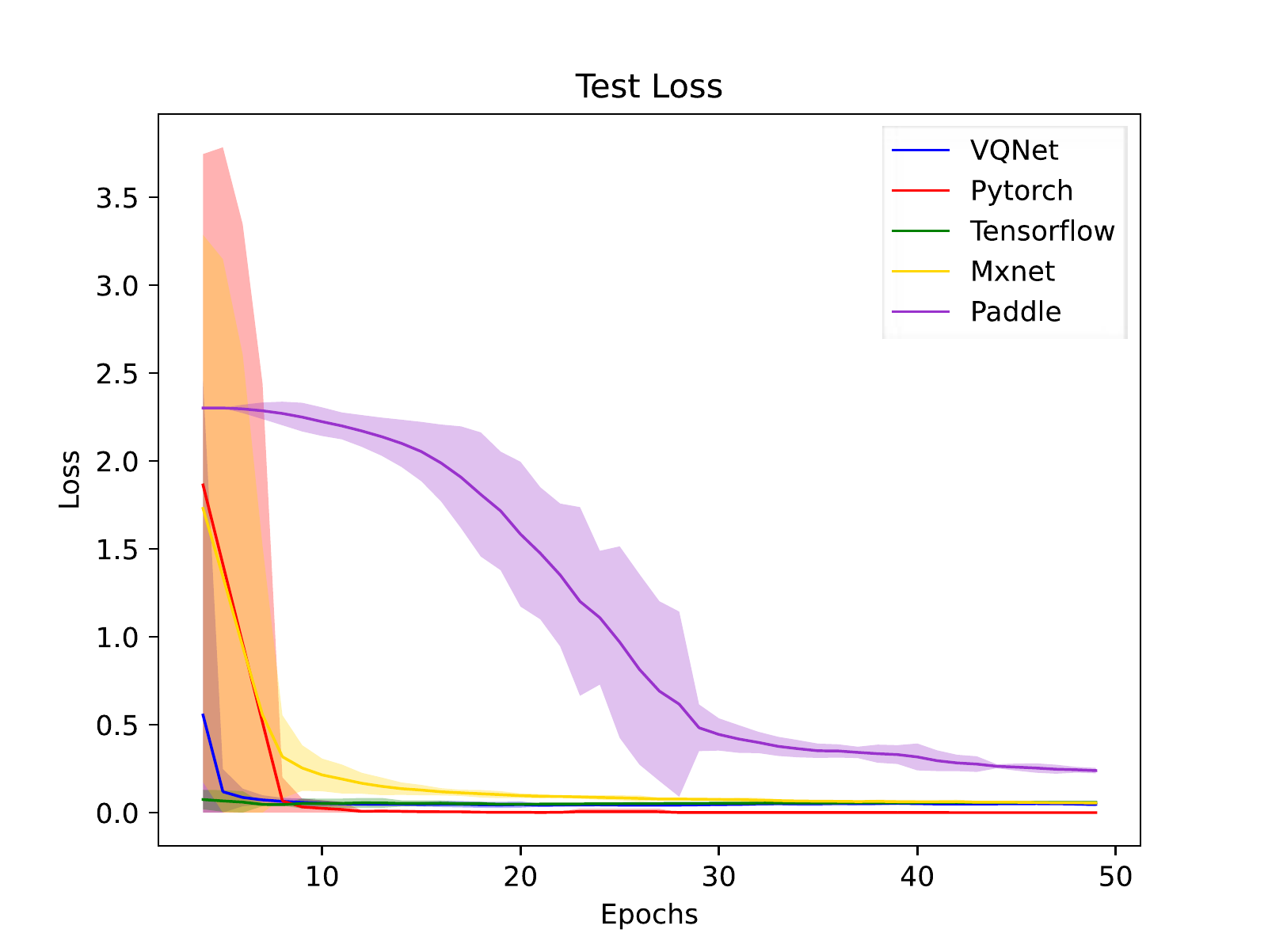}}\\
\subfloat[]{\includegraphics[width = 200pt]{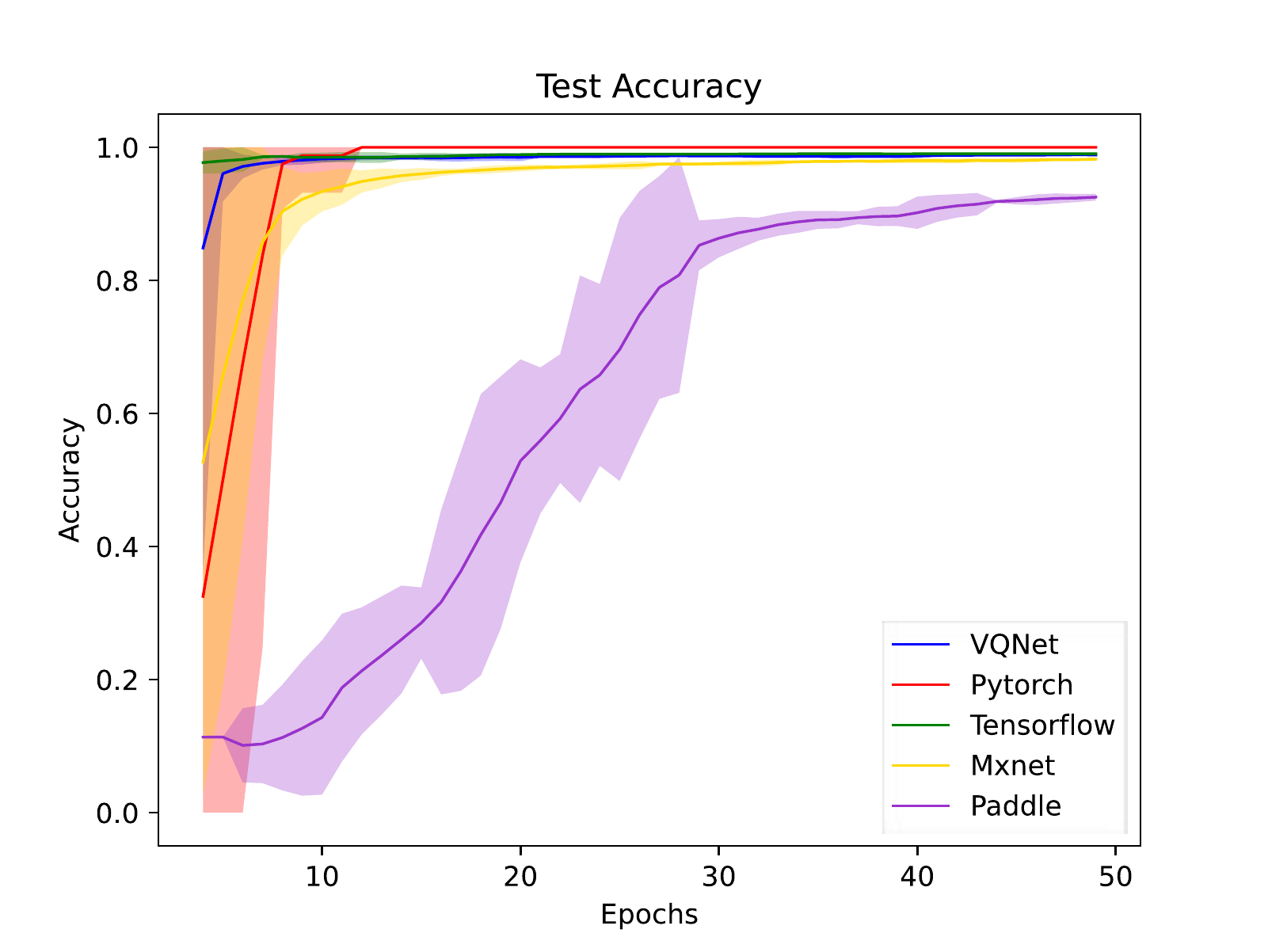}}\\
\caption{Loss and accuracy comparison of each machine learning framework in training and testing process.}
\label{subfigs}
\end{figure}

Firstly, we give the deployment results of the classical convolutional neural network implemented by VQNet 2.0 on the classical computer. The running instance is as described in Sec.\ref{sec:classic}. As shown in Fig.\ref{subfigs}, we can find that the classical convolutional neural network implemented by VQNet 2.0 converges after 10 epochs when training on the classical computer, and the test accuracy converges after the same number of epochs, which proves the deployment ability of VQNet 2.0 on the classical computer.

\begin{table*}[!htbp]
\scriptsize
\renewcommand\arraystretch{1.2}
\caption{Classical Machine Learning Deployment Loss}
\label{tab:different}
\begin{center}
\begin{tabular}{c|c|c|c|c|c}
\hline
      & Mxnet & Paddle & PyTorch & tensorflow & VQNet 2.0 \\
\hline\hline
LeNet  & 0.077 & 0.001 &\textbf{0.00} &0.002 &0.046  \\
\hline
AlexNet &0.050 &0.230 &\textbf{0.00} &0.059 &0.040\\
\hline
VGG  &0.022 &0.014 &\textbf{0.004} &0.176 &0.137\\
\hline
ResNet-18 &0.568 &0.112 &\textbf{0.082} &0.116 &0.118\\
\hline
\hline
\end{tabular}
\end{center}
\end{table*}

\begin{table*}[!htbp]
\scriptsize
\renewcommand\arraystretch{1.2}
\caption{Accuracy of Quantum Machine Learning Deployment}
\label{tab:different}
\begin{center}
\begin{tabular}{c|c|c|c|c|c}
\hline
      & Mxnet & Paddle & PyTorch & tensorflow & VQNet 2.0 \\
\hline\hline
LeNet   &0.975 &\textbf{1.00} &0.990  &0.989 &0.988\\
\hline
AlexNet &0.983 &0.927 &\textbf{1.00} &0.991  &0.999\\
\hline
VGG  &0.978 &0.976 &0.984 &0.98 &\textbf{0.997}\\
\hline
ResNet-18 &0.845 &\textbf{0.979} &0.973 &0.975 &0.971\\
\hline
\hline
\end{tabular}
\end{center}

\end{table*}

\begin{table*}[!htbp]
\scriptsize
\renewcommand\arraystretch{1.2}
\caption{Loss and Accuracy of Quantum Machine Learning Deployment on Quantum simulator}
\label{tab:different}
\begin{center}
\begin{tabular}{c|c|c|c|c}
\hline
      &Paddle Quantum & PyTorch+Pennylane & TensorFlow Quantum & VQNet 2.0 \\
\hline\hline
HQCNN &1.305/0.733 &1.682/\textbf{0.767} &1.726/\textbf{0.767}  &\textbf{1.209}/0.733\\
\hline
\hline
\end{tabular}
\end{center}

\end{table*}

\begin{table*}[!htbp]
\scriptsize
\renewcommand\arraystretch{1.2}
\caption{The duration of HQCNN is composed of different machine learning frameworks. BP stands for back propagation, FP stands for forward propagation. All data are in seconds.}
\label{tab:different}
\begin{center}
\begin{tabular}{c|c|c|c}
\hline
      & PyTorch + Qiskit & PyTorch + QPanda & VQNet 2.0 \\
\hline\hline
Total training duration & 62.55  & 19.75  & \textbf{12.91}  \\
\hline
 Single data training duration ($\times 10^{-3}$) & 15.38 & 4.66 & \textbf{3.36}  \\
\hline
 Quantum node BP duration($\times 10^{-4}$)  & 80.39 & 8.68 & \textbf{7.07}  \\
\hline
 Quantum node FP duration($\times 10^{-4}$)  & 41.83& 5.35 & \textbf{4.66} \\
\hline
 Network FP duration($\times 10^{-3}$)  & 4.92 & 1.29 & \textbf{1.04} \\
\hline
 Single data testing duration($\times 10^{-3}$) & 5.01 & 1.31 & \textbf{1.19}\\
\hline
 Dataset testing duration($\times 10^{-1}$) & 5.19 & 1.61 & \textbf{1.19}\\
\hline
\hline
\end{tabular}
\end{center}

\end{table*}

\subsubsection{Quantum Machine Learning Deployment}

This section shows the deployment results of VQNet 2.0 on the Wuyuan2 QPU which is a quantum computing hardware platform with 64 superconducting qubits created by Origin Quantum company based on Xmon architecture. According to whether there are quantum nodes or not, quantum machine learning can be divided into pure quantum networks and hybrid quantum-classical networks. Their running instances are shown in Sec.\ref{sec:qae} and Sec.\ref{sec:hqcnn} respectively. It is worth noting that in this section, the quantum nodes of the network operate on the quantum simulator, and the classical nodes are still deployed on the classical computer.

\subsection{Operation Efficiency}

By comparing our proposed VQNet 2.0 with the most popular quantum machine learning framework, we can fully understand the performance of VQNet 2.0. It is worth noting that since the VQNet 2.0 we proposed is the first unified framework for quantum and classical machine learning, the comparison methods in this section are the combination of the classical machine learning framework and quantum computing library. For example, PyTorch + Qiskit means using PyTorch to call a quantum circuit generated by Qiskit composes the completed HQCNN. For the fairness of comparison, we use different frameworks to train the same network model on the MNIST dataset: HQCNN, the loss function is cross-entropy loss, the optimizer is Adam, the learning rate reduction strategy is the same, and they are all trained for 20 epochs After stopping.

We first show the relationship between the loss of VQNet 2.0 and different comparison methods training the same network and the training epoch in Fig.\ref{subfigs}. Obviously, the loss of VQNet 2.0 decreases more smoothly and converges faster. As shown in Tab.\ref{tab:different}, the training and testing time of VQNet 2.0 is much shorter than the comparison method. For example, for the overall training time, PyTorch + Qiskit requires 62.5s, PyTorch+QPanda requires 19.7s, and VQNet 2.0 only requires 12.9s, which is 20\% of PyTorch + Qiskit, 65\% of PyTorch + QPanda. Thus, the efficient performance of our proposed framework is verified.

\subsection{Compatibility}
In order to show the compatibility of VQNet 2.0, we choose the quantum circuit built by Qiskit as a black box and send it to VQNet 2.0 for optimization. The running
instance is shown in Sec.\ref{sec:qiskit}.

\begin{figure}[!htbp]
\centering
\includegraphics[width = 220pt]{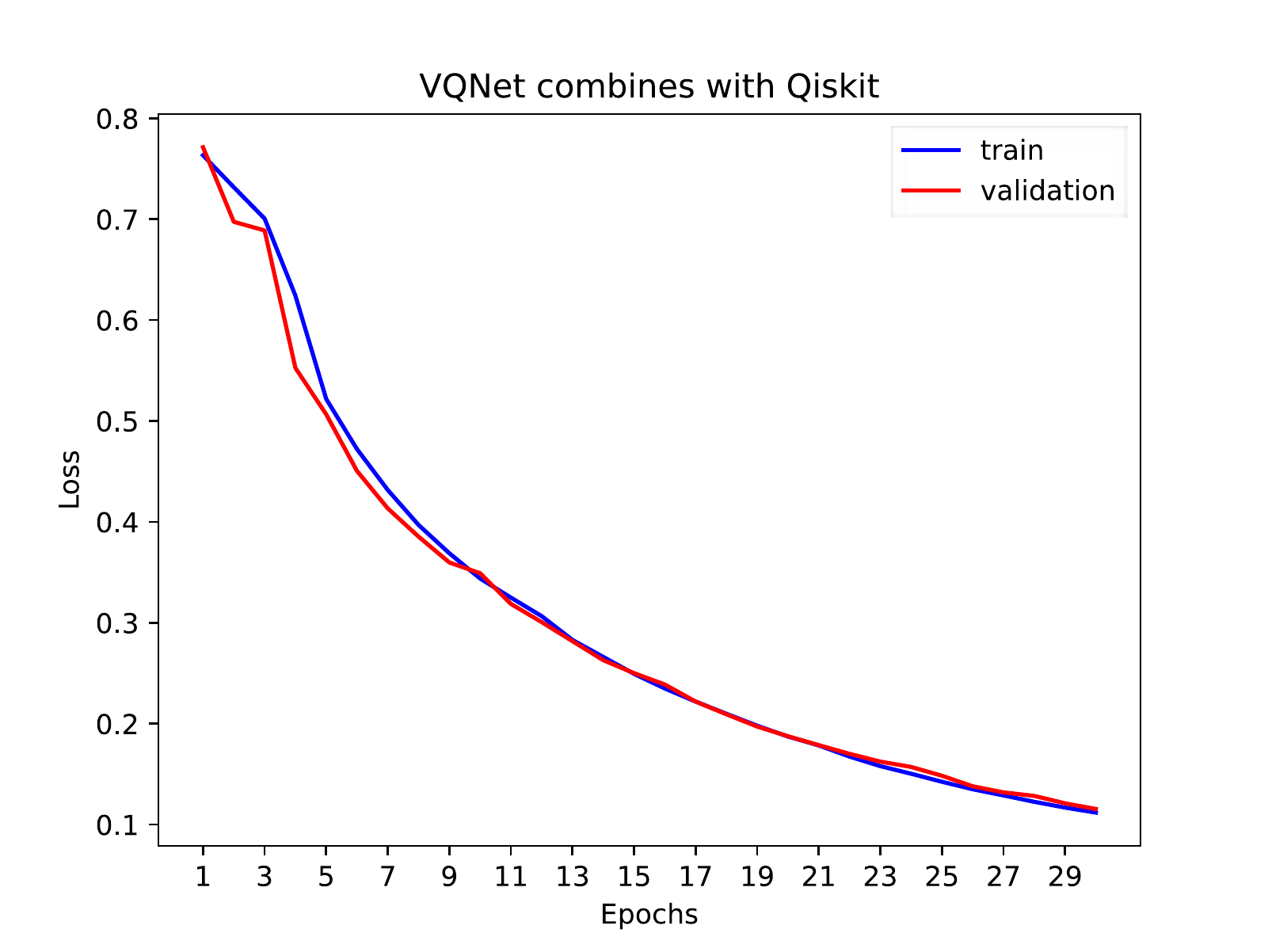}
\vspace{0.3em}
\caption{Loss of VQNet 2.0 uses the quantum circuit composed of Qiskit.}
\label{fig:qiskit}
\end{figure}

The relationship between training and testing loss is shown in Fig.\ref{fig:qiskit}. We can find that although VQNet 2.0 does not need to know the details of the quantum circuit composed by Qiskit, it can also carry out hybrid optimization and converge quickly. Experiments verify the compatibility of VQNet 2.0 with other quantum computing libraries.

\section{Conclusion and Future Work}
In this paper, on the basis of VQNet 1.0, we propose the first classical and quantum unified machine learning framework named VQNet 2.0. It supports both classical and quantum machine learning, and supports deployment on classical computers and QPU. VQNet 2.0 puts practicability in the first place through the design of the friendly interface, automatic differentiation, dynamic computation graph, and so on. Secondly, it provides superb performance by accomplishing a unified structure and quantum computing function based on QPanda. Finally, it can be used with other quantum computing libraries to provide better compatibility. Experiments prove that VQNet 2.0 can achieve better operating speed compared with the comparison method.

In the future, we plan to continue to improve the speed and practicality of VQNet 2.0 while supporting the latest developments in classical and quantum machine learning.
\section{Acknowledgments}
This work was supported in part by Anhui provincial major science and technology projects under Grant 202203a13010003, the National Natural Science Foundation of China under Grant 12034018, Innovation Program for Quantum Science and Technology under Grant 2021ZD0302300 and by the Natural Science Foundation of China under Grant 62272003.
\ifCLASSOPTIONcaptionsoff
  \newpage
\fi

\newpage
{\small
\bibliographystyle{IEEEtran}
\bibliography{egbib.bib}

\begin{thebibliography}{10}
\providecommand{\url}[1]{#1}
\csname url@samestyle\endcsname
\providecommand{\newblock}{\relax}
\providecommand{\bibinfo}[2]{#2}
\providecommand{\BIBentrySTDinterwordspacing}{\spaceskip=0pt\relax}
\providecommand{\BIBentryALTinterwordstretchfactor}{4}
\providecommand{\BIBentryALTinterwordspacing}{\spaceskip=\fontdimen2\font plus
\BIBentryALTinterwordstretchfactor\fontdimen3\font minus
  \fontdimen4\font\relax}
\providecommand{\BIBforeignlanguage}[2]{{%
\expandafter\ifx\csname l@#1\endcsname\relax
\typeout{** WARNING: IEEEtran.bst: No hyphenation pattern has been}%
\typeout{** loaded for the language `#1'. Using the pattern for}%
\typeout{** the default language instead.}%
\else
\language=\csname l@#1\endcsname
\fi
#2}}
\providecommand{\BIBdecl}{\relax}
\BIBdecl

\bibitem{huang2021power}
H.-Y. Huang, M.~Broughton, M.~Mohseni, R.~Babbush, S.~Boixo, H.~Neven, and
  J.~R. McClean, ``Power of data in quantum machine learning,'' \emph{Nature
  communications}, vol.~12, no.~1, pp. 1--9, 2021.

\bibitem{guan2021quantum}
W.~Guan, G.~Perdue, A.~Pesah, M.~Schuld, K.~Terashi, S.~Vallecorsa, and J.-R.
  Vlimant, ``Quantum machine learning in high energy physics,'' \emph{Machine
  Learning: Science and Technology}, vol.~2, no.~1, p. 011003, 2021.

\bibitem{lamata2020quantum}
L.~Lamata, ``Quantum machine learning and quantum biomimetics: A perspective,''
  \emph{Machine Learning: Science and Technology}, vol.~1, no.~3, p. 033002,
  2020.

\bibitem{gyongyosi2020optimizing}
L.~Gyongyosi and S.~Imre, ``Optimizing high-efficiency quantum memory with
  quantum machine learning for near-term quantum devices,'' \emph{Scientific
  reports}, vol.~10, no.~1, pp. 1--24, 2020.

\bibitem{biamonte2017quantum}
J.~Biamonte, P.~Wittek, N.~Pancotti, P.~Rebentrost, N.~Wiebe, and S.~Lloyd,
  ``Quantum machine learning,'' \emph{Nature}, vol. 549, no. 7671, pp.
  195--202, 2017.

\bibitem{ciliberto2018quantum}
C.~Ciliberto, M.~Herbster, A.~D. Ialongo, M.~Pontil, A.~Rocchetto, S.~Severini,
  and L.~Wossnig, ``Quantum machine learning: a classical perspective,''
  \emph{Proceedings of the Royal Society A: Mathematical, Physical and
  Engineering Sciences}, vol. 474, no. 2209, p. 20170551, 2018.

\bibitem{von2018quantum}
O.~A. Von~Lilienfeld, ``Quantum machine learning in chemical compound space,''
  \emph{Angewandte Chemie International Edition}, vol.~57, no.~16, pp.
  4164--4169, 2018.

\bibitem{christensen2020fchl}
A.~S. Christensen, L.~A. Bratholm, F.~A. Faber, and O.~Anatole~von Lilienfeld,
  ``Fchl revisited: Faster and more accurate quantum machine learning,''
  \emph{The Journal of chemical physics}, vol. 152, no.~4, p. 044107, 2020.

\bibitem{chen2019vqnet}
Z.-Y. Chen, C.~Xue, S.-M. Chen, and G.-P. Guo, ``Vqnet: Library for a
  quantum-classical hybrid neural network,'' \emph{arXiv preprint
  arXiv:1901.09133}, 2019.

\bibitem{menghan2022QPanda}
M.~Dou, T.~Zou, Y.~Fang, J.~Wang, D.~Zhao, L.~Yu, B.~Chen, W.~Guo, Y.~Li,
  Z.~Chen, and G.~Guo, ``Qpanda: high-performance quantum computing framework
  for multiple application scenarios,'' \emph{arXiv preprint arXiv:2212.14201},
  2022.

\bibitem{jia2014caffe}
Y.~Jia, E.~Shelhamer, J.~Donahue, S.~Karayev, J.~Long, R.~Girshick,
  S.~Guadarrama, and T.~Darrell, ``Caffe: Convolutional architecture for fast
  feature embedding,'' in \emph{Proceedings of the 22nd ACM international
  conference on Multimedia}, 2014, pp. 675--678.

\bibitem{abadi2016tensorflow}
M.~Abadi, P.~Barham, J.~Chen, Z.~Chen, A.~Davis, J.~Dean, M.~Devin,
  S.~Ghemawat, G.~Irving, M.~Isard \emph{et~al.}, ``Tensorflow: A system for
  large-scale machine learning,'' in \emph{12th $\{$USENIX$\}$ symposium on
  operating systems design and implementation ($\{$OSDI$\}$ 16)}, 2016, pp.
  265--283.

\bibitem{schreiber2017pomegranate}
J.~Schreiber, ``Pomegranate: fast and flexible probabilistic modeling in
  python,'' \emph{The Journal of Machine Learning Research}, vol.~18, no.~1,
  pp. 5992--5997, 2017.

\bibitem{paszke2019pytorch}
A.~Paszke, S.~Gross, F.~Massa, A.~Lerer, J.~Bradbury, G.~Chanan, T.~Killeen,
  Z.~Lin, N.~Gimelshein, L.~Antiga \emph{et~al.}, ``Pytorch: An imperative
  style, high-performance deep learning library,'' \emph{Advances in neural
  information processing systems}, vol.~32, pp. 8026--8037, 2019.

\bibitem{broughton2020tensorflow}
M.~Broughton, G.~Verdon, T.~McCourt, A.~J. Martinez, J.~H. Yoo, S.~V. Isakov,
  P.~Massey, R.~Halavati, M.~Y. Niu, A.~Zlokapa \emph{et~al.}, ``Tensorflow
  quantum: A software framework for quantum machine learning,'' \emph{arXiv
  preprint arXiv:2003.02989}, 2020.

\bibitem{bergholm2018pennylane}
V.~Bergholm, J.~Izaac, M.~Schuld, C.~Gogolin, M.~S. Alam, S.~Ahmed, J.~M.
  Arrazola, C.~Blank, A.~Delgado, S.~Jahangiri \emph{et~al.}, ``Pennylane:
  Automatic differentiation of hybrid quantum-classical computations,''
  \emph{arXiv preprint arXiv:1811.04968}, 2018.

\bibitem{Paddlequantum}
``{Paddle Quantum},'' \emph{https://github.com/PaddlePaddle/Quantum}, 2020.

\bibitem{baydin2018automatic}
A.~G. Baydin, B.~A. Pearlmutter, A.~A. Radul, and J.~M. Siskind, ``Automatic
  differentiation in machine learning: a survey,'' \emph{Journal of machine
  learning research}, vol.~18, 2018.

\bibitem{paszke2017automatic}
A.~Paszke, S.~Gross, S.~Chintala, G.~Chanan, E.~Yang, Z.~DeVito, Z.~Lin,
  A.~Desmaison, L.~Antiga, and A.~Lerer, ``Automatic differentiation in
  pytorch,'' 2017.

\bibitem{van1995python}
G.~Van~Rossum and F.~L. Drake~Jr, \emph{Python tutorial}.\hskip 1em plus 0.5em
  minus 0.4em\relax Centrum voor Wiskunde en Informatica Amsterdam, 1995, vol.
  620.

\bibitem{oliphant2007python}
T.~E. Oliphant, ``Python for scientific computing,'' \emph{Computing in science
  \& engineering}, vol.~9, no.~3, pp. 10--20, 2007.

\bibitem{meyers2005effective}
S.~Meyers, \emph{Effective C++: 55 specific ways to improve your programs and
  designs}.\hskip 1em plus 0.5em minus 0.4em\relax Pearson Education, 2005.

\bibitem{alexandrescu2001modern}
A.~Alexandrescu, \emph{Modern C++ design: generic programming and design
  patterns applied}.\hskip 1em plus 0.5em minus 0.4em\relax Addison-Wesley,
  2001.

\bibitem{collobert2011torch7}
R.~Collobert, K.~Kavukcuoglu, and C.~Farabet, ``Torch7: A matlab-like
  environment for machine learning,'' in \emph{BigLearn, NIPS workshop}, no.
  CONF, 2011.

\bibitem{chetlur2014cudnn}
S.~Chetlur, C.~Woolley, P.~Vandermersch, J.~Cohen, J.~Tran, B.~Catanzaro, and
  E.~Shelhamer, ``cudnn: Efficient primitives for deep learning,'' \emph{arXiv
  preprint arXiv:1410.0759}, 2014.

\bibitem{lavin2015maxdnn}
A.~Lavin, ``maxdnn: an efficient convolution kernel for deep learning with
  maxwell gpus,'' \emph{arXiv preprint arXiv:1501.06633}, 2015.

\bibitem{Mindspore}
\BIBentryALTinterwordspacing
H.~M. team, ``Mindspore,'' 2019. [Online]. Available:
  \url{https://github.com/mindspore-ai/mindspore}
\BIBentrySTDinterwordspacing

\end{thebibliography}
}

\newpage
\appendix[Running Instances]

\subsection{Classical Convolutional Neural Network for Image Recognition}
\label{sec:classic}
\subsubsection{Background}

Neural networks are part of artificial intelligence research. The most popular neural networks are convolutional neural networks(CNN). CNNs have achieved great success in many research fields, such as speech recognition, image recognition, image segmentation, natural language processing. Convolutional neural networks are mainly composed of these types of layers: input layer, convolutional layer, ReLU layer, pooling layer, and fully connected layer, etc.
\subsubsection{Implementations}
As mentioned earlier, the first step of the VQNet 2.0 process is data set preprocessing. In this example, we use the common MNIST data set in computer vision as the experimental data set. The MNIST data set is the handwritten character library. There are 70,000 images, of which 60,000 are the training set and 10,000 are the test set:

\begin{lstlisting}
def load_mnist(dataset="training_data", digits=np.arange(2), path="..//..//data//MNIST_data"): 
    import os, struct
    from array import array as pyarray
    if dataset == "training_data":
        fname_image = os.path.join(path, 'train-images.idx3-ubyte').replace('\\', '/')
        fname_label = os.path.join(path, 'train-labels.idx1-ubyte').replace('\\', '/')
    elif dataset == "testing_data":
        fname_image = os.path.join(path, 't10k-images.idx3-ubyte').replace('\\', '/')
        fname_label = os.path.join(path, 't10k-labels.idx1-ubyte').replace('\\', '/')
    else:
        raise ValueError("dataset must be 'training_data' or 'testing_data'")
\end{lstlisting}

Once the dataset is ready, we start building the model. The model consists of two convolution layers and one full connection layer. Each convolution layer performs the max-pooling operation after the relu activation function.

\begin{lstlisting}
class CNN(Module):
    def __init__(self):
        super(CNN, self).__init__()
        self.maxpool4 = MaxPool2D([2, 2], [2, 2], padding="valid")
        self.maxpool2 = MaxPool2D([2, 2], [2, 2], padding="valid")
        self.conv5 = Conv2D(input_channels=1, output_channels=32, kernel_size=(3, 3), stride=(1, 1), padding="valid")
        self.conv6 = Conv2D(input_channels=32, output_channels=32, kernel_size=(3, 3), stride=(1, 1), padding="valid")
        self.fc3 = Linear(input_channels=800, output_channels=10)
        self.Relu2 = F.ReLu()

    def forward(self, x):
        x = self.maxpool2(self.Relu2(self.conv5(x)))
        x = self.maxpool4(self.Relu2(self.conv6(x)))
        x = tensor.flatten(x,1)
        x = self.fc3(x)

        return x
\end{lstlisting}

In this example, we set the loss function as cross-entropy loss, and the optimizer is a common Adam optimization. If we set the maximum number of epochs and batch size, we can start training the whole model.
\begin{lstlisting}
model = CNN()
model.train()

loss_func = SoftmaxCrossEntropy()

model_hybrid = model

optimizer_hybrid = Adam(model_hybrid.parameters(), lr=0.001)

for epoch in range(1, epochs):
    total_loss = []
    train_acc = 0
    for x, y in data_generator(x_train, y_train, batch_size=16, shuffle=True):
        x = x.reshape(-1, 1, 28, 28)
        optimizer_hybrid.zero_grad()

        # Forward pass
        output = model_hybrid(x)
        np_output = np.array(output.data)
        loss = loss_func(y, output)
        loss_np = np.array(loss.data)

        # Backward pass
        loss.backward()
        optimizer_hybrid._step()
        total_loss.append(loss_np)
\end{lstlisting}
\subsection{Quantum Autoencoder for Efficient Compression of Quantum Data}
\label{sec:qae}

\subsubsection{Background}
Inspired by the classical autoencoder, the model of the quantum autoencoder is used to perform similar tasks on quantum data. Quantum autoencoder is trained to compress specific data sets of quantum states, while classical compression algorithms cannot be used. The parameters of the quantum autoencoder are trained by the classical optimization algorithm. We show an example of a simple programmable circuit, which can be trained into an efficient autoencoder. In the context of quantum simulation, we apply our model to compress the ground state of the Hubbard model and molecular Hamiltonian.

\subsubsection{Implementations}
We first define \Colorbox{bkgd}{\lstinline{QAElayer}}, which is parameterized quantum circuit Layer. It contains parameters that can be trained.
\begin{lstlisting}
class QAElayer(Module):
    def __init__(self,trash_qubits_number:int=2,total_qubits_number:int=7,machine:str="cpu"):
        super().__init__()

        self.machine = machine
        if machine!="cpu":
            raise ValueError("machine only tested on cpu simulation")
        self.machine = pq.CPUQVM()
        self.machine.init_qvm()
        self.qlist = self.machine.qAlloc_many(total_qubits_number)
        aux_qubits_number = 1
        self.clist = self.machine.cAlloc_many(aux_qubits_number)

        self.history_prob = []

        self.n_qubits = total_qubits_number
        self.n_aux_qubits = aux_qubits_number
        self.n_trash_qubits = trash_qubits_number

        training_qubits_size = self.n_qubits - self.n_aux_qubits - self.n_trash_qubits

        weight_shapes = {"params_rot_begin": training_qubits_size * 3,
                     "params_crot": training_qubits_size * (training_qubits_size - 1) * 3,
                     "params_rot_end":  training_qubits_size * 3}

        self.weights = Parameter(weight_shapes['params_rot_begin']
                                + weight_shapes['params_crot']
                                + weight_shapes['params_rot_end'])

    def forward(self, x):
        self.history_prob = []
        batchsize = x.shape[0]
        batch_measure = np.zeros([batchsize])

        prog = pqc(x.data,
                   self.weights.data,
                   self.qlist,
                   self.clist,
                   self.n_qubits,
                   self.n_aux_qubits,
                   self.n_trash_qubits)

        result = self.machine.run_with_configuration(prog, self.clist, 100)
        counts = result['0']
        probabilities = counts / 100
        print("probabilities", probabilities)
        batch_measure[0] = probabilities

        requires_grad = (x.requires_grad or self.weights.requires_grad) and not QTensor.NO_GRAD
        nodes = []
        if x.requires_grad:
            nodes.append(QTensor.GraphNode(tensor=x, df=lambda g: 1))
        if self.weights.requires_grad:
            nodes.append(QTensor.GraphNode(
            tensor = self.weights, 
            df=lambda g: _grad(g,
                              pqc,
                              x.data,
                              self.weights.data,
                              self.machine,
                              self.qlist,self.clist,
                              self.n_qubits,
                              self.n_aux_qubits,
                              self.n_trash_qubits
                              )))
        return QTensor(data = [batch_measure],requires_grad = requires_grad,nodes =nodes)
\end{lstlisting}

Finally, the forward of the whole network is set, and the network training can be conducted with common training modules:
\begin{lstlisting}
class Model(Module):
    def __init__(self, trash_num: int = 2, total_num: int = 7):
        super().__init__()
        self.pqc = QAElayer(trash_num, total_num)

    def forward(self, x):
        x = self.pqc(x)

        return x
\end{lstlisting}

\subsection{Hybrid Quantum-Classical Neural Networks in VQNet 2.0}
\label{sec:hqcnn}
\subsubsection{Background}
We explore how a classical neural network can be partially quantized to create a hybrid quantum-classical neural network. We will code up a simple example that integrates QPanda with VQNet 2.0.

\subsubsection{Implementations}

We create a simple hybrid neural network from MNIST datasets, and two types of digital (0 or 1) images are classified. We first load MNIST and pick out the images containing 0 and 1. These are used as the input of the neural network for classification.
\begin{lstlisting}
def load_mnist(dataset="training_data", digits=np.arange(2), path="../../../dataset/MNIST_data"):        
    import os, struct
    from array import array as pyarray
    if dataset == "training_data":
        fname_image = os.path.join(path, 'train-images.idx3-ubyte').replace('\\', '/')
        fname_label = os.path.join(path, 'train-labels.idx1-ubyte').replace('\\', '/')
    elif dataset == "testing_data":
        fname_image = os.path.join(path, 't10k-images.idx3-ubyte').replace('\\', '/')
        fname_label = os.path.join(path, 't10k-labels.idx1-ubyte').replace('\\', '/')
    else:
        raise ValueError("dataset must be 'training_data' or 'testing_data'")

    flbl = open(fname_label, 'rb')
    magic_nr, size = struct.unpack(">II", flbl.read(8))
    lbl = pyarray("b", flbl.read())
    flbl.close()

    fimg = open(fname_image, 'rb')
    magic_nr, size, rows, cols = struct.unpack(">IIII", fimg.read(16))
    img = pyarray("B", fimg.read())
    fimg.close()

    ind = [k for k in range(size) if lbl[k] in digits]
    N = len(ind)
    images = np.zeros((N, rows, cols))
    labels = np.zeros((N, 1), dtype=int)
    for i in range(len(ind)):
        images[i] = np.array(img[ind[i] * rows * cols: (ind[i] + 1) * rows * cols]).reshape((rows, cols))
        labels[i] = lbl[ind[i]]

    return images, labels
\end{lstlisting}

This experiment is a binary classification task, so one qubit is used.

\begin{lstlisting}
def circuit(weights):
    num_qubits = 1
    machine = pq.CPUQVM()
    machine.init_qvm()
    qubits = machine.qAlloc_many(num_qubits)
    cbits = machine.cAlloc_many(num_qubits)
    circuit = pq.QCircuit()

    circuit.insert(pq.H(qubits[0]))
    circuit.insert(pq.RY(qubits[0], weights[0]))

    prog = pq.QProg()
    prog.insert(circuit)
    prog << measure_all(qubits, cbits)

    result = machine.run_with_configuration(prog, cbits, 100)
    counts = np.array(list(result.values()))
    states = np.array(list(result.keys())).astype(float)
    # Compute probabilities for each state
    probabilities = counts / 100
    # Get state expectation
    expectation = np.sum(states * probabilities)
    return expectation
\end{lstlisting}

Here, we first reduce the data's dimension to 2D through the pooling layer and the full connection layer, then through the single qubit circuit, and finally into 2D output through the full connection layer to adapt to the 2-classification task.
\begin{lstlisting}
class Hybrid(Module):
    def __init__(self, shift):
        super(Hybrid, self).__init__()
        self.shift = shift

    def forward(self, input):
        self.input = input
        expectation_z = circuit(np.array(input.data))
        result = [[expectation_z]]
        requires_grad = input.requires_grad and not QTensor.NO_GRAD

        def _backward_mnist(g, input):
            input_list = np.array(input.data)
            shift_right = input_list + np.ones(input_list.shape) * self.shift
            shift_left = input_list - np.ones(input_list.shape) * self.shift

            gradients = []
            for i in range(len(input_list)):
                expectation_right = circuit(shift_right[i])
                expectation_left = circuit(shift_left[i])

                gradient = expectation_right - expectation_left
                gradients.append(gradient)
            gradients = np.array([gradients]).T
            return gradients * np.array(g)

        nodes = []
        if input.requires_grad:
            nodes.append(QTensor.ComputationalGraphNode(tensor=input, df=lambda g: _backward_mnist(g, input)))

        return QTensor(data=result, requires_grad=requires_grad, nodes=nodes)

class Net(Module):
    def __init__(self):
        super(Net, self).__init__()
        self.conv1 = Conv2D(input_channels=1, output_channels=6, kernel_size=(5, 5), stride=(1, 1), padding="valid")
        self.maxpool1 = MaxPool2D([2, 2], [2, 2], padding="valid")
        self.conv2 = Conv2D(input_channels=6, output_channels=16, kernel_size=(5, 5), stride=(1, 1), padding="valid")
        self.maxpool2 = MaxPool2D([2, 2], [2, 2], padding="valid")



        self.fc1 = Linear(input_channels=256, output_channels=64)
        self.fc2 = Linear(input_channels=64, output_channels=1)

        self.hybrid = Hybrid(np.pi / 2)
        self.fc3 = Linear(input_channels=1, output_channels=2)

    def forward(self, x):
        x = F.ReLu()(self.conv1(x))

        x = self.maxpool1(x)
        x = F.ReLu()(self.conv2(x))

        x = self.maxpool2(x)
        x = tensor.flatten(x, 1)

        x = F.ReLu()(self.fc1(x))
        x = self.fc2(x) 
        x = self.hybrid(x)

        x = self.fc3(x)
        return x
\end{lstlisting} 

\subsection{Using Qiskit in VQNet 2.0}
\label{sec:qiskit}
\subsubsection{Background}
In VQNet 2.0, we can also use IBM's Qiskit quantum computing library for quantum machine learning tasks.

VQNet 2.0 implements the automatic differential Qiskit quantum circuit operation class \Colorbox{bkgd}{\lstinline{QiskitLayer}}, which inherits from \Colorbox{bkgd}{\lstinline{Compatiblelayer}}. \Colorbox{bkgd}{\lstinline{Compatiblelayer}} is a class that is compatible with other framework circuits to VQNet 2.0. The parameters of constructing \Colorbox{bkgd}{\lstinline{QiskitLayer}} need to be passed in a class, which defines the Qiskit quantum circuit \Colorbox{bkgd}{\lstinline{qiskit.QuantumCircuit}}, as well as its operation and measurement function \Colorbox{bkgd}{\lstinline{run}}. The \Colorbox{bkgd}{\lstinline{run}} function needs to bind the input value to the quantum circuit of Qiskit. Using \Colorbox{bkgd}{\lstinline{QiskitLayer}}, the input of quantum circuits and automatic differentiation can be implemented by VQNet 2.0.

\subsubsection{Implementations}

We use Qiskit to construct a class \Colorbox{bkgd}{\lstinline{QISKI_VQC}}. The variable \Colorbox{bkgd}{\lstinline{self._circuit}} is a quantum circuit. \Colorbox{bkgd}{\lstinline{self.input}} is a variable input parameter. In the run function, you need to use Qiskit's \Colorbox{bkgd}{\lstinline{assign_parameters}} to bind the parameters and use \Colorbox{bkgd}{\lstinline{self.backend.run}} to run.

\begin{lstlisting}
class QISKIT_VQC:
        def __init__(self, n_qubits, backend, shots):
                # --- Circuit definition ---
                self._circuit = qiskit.QuantumCircuit(n_qubits)

                all_qubits = [i for i in range(n_qubits)]
                self.input = [qiskit.circuit.Parameter('input')]

                self._circuit.h(all_qubits)
                self._circuit.barrier()
                self._circuit.ry(self.input[0], all_qubits)

                self._circuit.measure_all()

                self.backend = backend
                self.shots = shots

        def run(self,x):
                params = dict(zip(self.input, x))
                c1 = self._circuit.assign_parameters(params)

                job = self.backend.run(c1,shots=self.shots)
                result = job.result().get_counts()

                counts = np.array(list(result.values()))
                states = np.array(list(result.keys())).astype(float)

                # Compute probabilities for each state
                probabilities = counts / self.shots
                # Get state expectation
                expectation = np.sum(states * probabilities)

                return expectation
\end{lstlisting} 

The next step is to use VQNet 2.0 to define the model and training process. Use \Colorbox{bkgd}{\lstinline{QiskitLayer}} to add the Qiskit circuit to the model.
\begin{lstlisting}
#define Qiskit circuits class
circuit = QISKIT_VQC(1, simulator, 100)

class Net(Module):
        def __init__(self):
                super(Net, self).__init__()
                self.conv1 = Conv2D(input_channels=1, output_channels=6, kernel_size=(5, 5), stride=(1, 1), padding="valid")
                self.maxpool1 = MaxPool2D([2, 2], [2, 2], padding="valid")
                self.conv2 = Conv2D(input_channels=6, output_channels=16, kernel_size=(5, 5), stride=(1, 1), padding="valid")
                self.maxpool2 = MaxPool2D([2, 2], [2, 2], padding="valid")
                self.fc1 = Linear(input_channels=256, output_channels=64)
                self.fc2 = Linear(input_channels=64, output_channels=1)
                self.hybrid = QiskitLayer(circuit,0)
                self.fc3 = Linear(input_channels=1, output_channels=2)

        def forward(self, x):
                x = F.ReLu()(self.conv1(x))
                x = self.maxpool1(x)
                x = F.ReLu()(self.conv2(x))
                x = self.maxpool2(x)
                x = tensor.flatten(x, 1)
                x = F.ReLu()(self.fc1(x))
                x = self.fc2(x) 
                x = self.hybrid(x)
                x = self.fc3(x)
                return x
\end{lstlisting} 

\end{document}